\documentclass[12pt,amsmath,amssymb,showpacs]{revtex4}
\newcommand {\eq}{\begin{equation}}
\newcommand {\qe}{\end{equation}}
\newcommand {\cen}[1]{\begin{center} #1 \end{center}}
\newcommand {\bfr}{{\bf r}}
\newcommand {\bfq}{{\bf q}}
\newcommand {\bfs}{{\bf s}}
\newcommand {\bfb}{{\bf b}}
\newcommand {\bfc}{{\bf c}}
\newcommand {\bfd}{{\bf d}}
\newcommand {\bfa}{{\bf a}}
\newcommand {\bft}{{\bf t}}
\newcommand {\bfu}{{\bf u}}
\newcommand {\bfv}{{\bf v}}
\newcommand {\bfx}{{\bf x}}
\newcommand {\bfz}{{\bf z}}
\newcommand {\bfR}{{\bf R}}
\newcommand {\bfQ}{{\bf Q}}
\newcommand {\Gc}{{\rm GeV/c}}
\newcommand {\ea} {{\it et al.}}

\newcommand {\bfk}{{\bf k}}
\newcommand {\bfp}{{\bf p}}
\newcommand {\h}{\frac{1}{2}}

\newcommand {\bfkap}{\mbox {\boldmath $\kappa$}}

\newcommand {\tal}{\tilde{\alpha}}
\newcommand {\tbe}{\tilde{\beta}}

\newcommand {\np}{Nucl. Phys. }
\newcommand {\nucp}{Nucl. Phys. }
\newcommand {\bfP}{{\bf P}}

\usepackage{graphicx}% Include figure files
\usepackage{epsfig}
\begin{document}
\today

%\vspace*{-1.0in}

%\hspace*{3.5in}{\Huge \bf Rough Draft}

%\hspace*{4in}\today

%\vspace*{1in}

\title{K$^+$-nucleon interaction}

\author{ W. R. Gibbs and R. Arceo}
\affiliation{ Department of Physics, New Mexico State University 
\\
 Las Cruces, New Mexico 88003, USA\\}

\begin{abstract}

\cen{\bf Abstract} 

The Born approximation to the quark-gluon-exchange mechanism for K$^+$N 
scattering is used as a starting point to generate a potential for this 
system. The valence quark wave function of the nucleon is generalized from 
a single Gaussian to a sum of Gaussians in order to have a more flexible 
representation than previous work. We obtain a potential derived from a 
valence density given by lattice calculations. By comparing with a recent 
amplitude analysis it is found that the strength of the quark-gluon based 
potential needs to be increased by a factor of order 2-4 relative to the 
normalization given by more traditional values of the governing 
parameters. The method is used to estimate the change in effective K$^+$N 
amplitudes which would result from changes in the valence quark 
distributions or strength of the interaction which might arise from 
nuclear medium effects in K$^+$ scattering from nuclei.

\end{abstract}


\maketitle

\newpage

\section{Introduction}

The K$^+$ meson is useful in the study of the strong interaction
because of its simple structure, i.e. it has no light valence 
antiquark. This feature means that no 3-quark intermediate state
can exist, thus removing the most common resonances from the 
interaction. Although the interaction is weaker than other typical
strong interactions it is still substantial and must be treated as 
such.

One of the principal motivations of this work is to obtain the 
relationship between the distribution of the constituents of the nucleon  
and the $K^+$-N scattering amplitudes.
In Ref. \cite{siegel} it was suggested that the increase in size of
the nucleon in the nucleus could be observed as an alteration in the
scattering of the kaons from nuclei. In that case the model used to
connect the change in size with scattering was a strongly repulsive
well whose radius was allowed to vary. While this simple model may
be expected to encompass much of the basic physics, clearly a model
based on fundamental degrees of freedom is to be preferred.

Barnes and Swanson \cite{bs} treated the K$^+$-nucleon interaction due 
to the exchange of one gluon and two quarks, one of the lowest order 
process dealing directly with the underlying degrees of freedom. (Barnes, 
Black and Swanson\cite{bands2} carried out a similar program for 
meson-meson scattering but we do not address that sector here). Their 
calculation of a lowest order exchange must be iterated to obtain a full 
evaluation of the scattering amplitude. A common way to do this is to 
treat their result as the Born approximation corresponding to a potential 
and use a wave equation to generate an approximation to the full 
amplitude.

Such a program has been very successful in the case of one pion 
exchange. The one-pion-exchange potential is able to predict many 
properties of the deuteron \cite{rosa,fg,ballot}. In the pion case 
the exchange of a single particle gives rise to a local potential. 
In the present (quark-gluon exchange) case the potential has one 
term which is local and three others which are non-local.

In the case of one pion exchange, a potential is essential in 
order to be able to produce the deuteron bound state. In the 
present work the potential is needed to calculate realistic phase 
shifts for comparison with data.  This potential will then enable 
us to relate possible changes in valence quark distribution or 
strengths to the changes in phase shifts which would result.

Because they included no mechanism that would lead to spin 
dependence, the results of Ref. \cite{bs} were necessarily 
spin-independent. The experimentally determined partial waves with 
$\ell$ greater than zero show a strong spin dependence\cite{ga} so 
that a comparison with these partial waves is not possible and 
they concentrated their work on the $\ell=0$ partial wave. We will 
do the same.

In a later, similar, program N. Black\cite{black} calculated the 
higher partial waves with quark-gluon exchange including the spin 
dependence.  For $T$=1 he found qualitative agreement except for 
the $P_{3/2}$ partial wave where the sign was wrong. For $T=0$ the 
phases shifts were considerably smaller than the experimental 
ones, although there was a tendency for an alternation of sign 
similar to what is observed. In any case, the $\ell=0$ partial 
wave dominates the isospin one amplitude at low to moderate 
energies.

The Barnes and Swanson\cite{bs} calculation assumes a Gaussian 
wave function for the distribution of the quarks in the nucleon. 
This is a rather restrictive assumption on the shape and does not 
represent the functional form found in lattice 
calculations\cite{lissia,alex}. To repeat their calculation 
directly with more realistic wave functions would be difficult 
because of the 9-dimensional integrals involved. With Gaussian 
wave functions these integrals can be done analytically so that an 
alternative method of treating this problem is by representing the 
desired wave function as a sum of Gaussian functions.  In this 
case the integrals can be done term by term but involve cross 
terms in which the Gaussian parameters in the initial and final 
states are different. Such an approach requires a derivation of 
the expressions of Ref. \cite{bs} for this more general condition.  
We carry out this step and then calculate a potential for a wave 
function consisting of the sum of two Gaussian functions.  One 
could include more terms but at this stage the knowledge of the 
desired wave function is probably not sufficiently accurate to 
warrant the effort.

In the calculation of Barnes and Swanson\cite{bs} the Born 
approximation to the amplitude for K$^+$N scattering is given by
\eq
A(\bfk,\bfk')=\frac{2\gamma m_R}{3}
\sum_{i=1}^4 \eta_iw_i F_i(\bfk,\bfk')
\qe
where
\eq
 F_i(\bfk,\bfk')=e^{-a_i k^2+b_i \bfk\cdot \bfk'}, \ \ \ 
m_R=\frac{E_KE_N}{E_K+E_N}, 
\qe 
$\bfk$ and $\bfk'$ are center of mass momenta, $k=|\bfk|=|\bfk'|$, 
$\gamma\equiv\alpha_s/m_q^2$, $E_N$ is the total energy of the 
nucleon and $E_K$ is the total energy of the kaon in the center of 
mass, $\alpha_s$ is the running strong coupling constant and $m_q$ 
is the light (constituent) quark mass. Expressions for the $a_i$, 
$b_i$, and $\eta_i$ are given in Ref. \cite{bs} for simple 
harmonic oscillator wave functions. They are functions of $\rho$, 
the ratio of light to strange quark (taken to be 0.6), $\alpha$, 
the parameter governing the size of the nucleon and $\beta$, the 
parameter determining the size of the kaon through the ratio 
$g=\alpha^2/\beta^2$.

They considered four diagrams for quark and gluon exchange.  For 
the first diagram the momentum transfer is given by the two 
exchanged quarks with gluon exchange between them so the resulting 
interaction is local as indicated by the fact that $a_1=b_1$. For 
the other three diagrams this is not the case and the interaction 
is non-local. Barnes and Swanson calculated an equivalent local 
potential valid at zero energy only.

The isospin dependence is contained in the weights, $w_i$, which 
are given by
\eq
\{w\}=\left\{0,\frac{1}{6},0,\frac{1}{6}\right\} :T=0{\rm \ 
and}\ \ 
\{w\}=\left\{\frac{1}{3},\frac{1}{18},\frac{1}{3},\frac{1}{18}\right\}
:T=1. \label{isofacs}
\qe
One sees that there is a bias toward the isospin unity amplitude being 
larger than isospin zero. If the contributions from each of the
four diagrams were equal (they need not be, of course, and are not) then
the $T=1$ amplitude would be 7/3 of the $T=0$ amplitude.

It is interesting to examine the low-energy limit of this amplitude. In 
this case the functions, $ F_i(\bfk,\bfk')$, become unity. The functions,
$\eta_i$ depend only on $g$ so, in this limit, the absolute size of the 
kaon and nucleon systems does not enter but their {\it relative} size 
remains very important.

In Section II we derive expressions for the potential for arbitrary values 
of the parameters in each term, in Section III we study the results with 
the use of various single Gaussian wave functions for the valence quarks, 
and in Section IV we develop the two-Gaussian form for the wave function. 
In Section V we give a summary of the formulas needed to calculate the 
parameters for the two-(or multi-)Gaussian potential, in Section VI we 
show the basic fit with the two-Gaussian form and study the consequences 
of a variation of the radius or strength (as for the case of immersion in 
nuclear matter). Section VII gives the results of a calculation of the 
off-shell amplitude which results from the two-Gaussian form for the 
potential and Section VIII gives some conclusions.

\section{Definition of the potential}

The potential will be given in terms of the amplitude by
\eq
V(\bfr,\bfr')=\frac{2\pi}{m_R}\frac{2\gamma m_R}{3}
\sum_{i=1}^4 \eta_iw_i F_i(\bfr,\bfr')
\qe
where
\eq
F_i(\bfr,\bfr')=\frac{1}{(2\pi)^6}\int d\bfk d\bfk' 
e^{i\bfk\cdot\bfr-i\bfk'\cdot\bfr'}F_i(\bfk,\bfk').
\qe

To get a fully off-shell amplitude which reduces to
this form on shell but is a function only of momentum
transfer if $a=b$, we write (dropping for the moment the index $i$)

\eq
F(\bfk,\bfk')=e^{-\frac{a}{2}(k^2+k'^2)+\bfk\cdot\bfk'b}
\qe
where $\bfk$ and $\bfk'$ are independent and are assumed not to be 
constrained by the on-shell condition. We hasten to point out that 
this is a choice and that other dependences are possible. We make 
no claim to uniqueness. We now choose new vector variables as
\eq
\bfkap\equiv \frac{\bfk+\bfk'}{2};\ \ 
\bfq\equiv\bfk-\bfk' \ \ {\rm so\ that}\ \ 
\bfk=\bfkap+\frac{\bfq}{2};\ \ \ \bfk'=\bfkap-
\frac{\bfq}{2}
\qe
allowing us to express the form as
\eq
F(\bfkap,\bfq)=e^{-\kappa^2(a-b)}e^{-q^2(a+b)/4}.
\qe

For the special case of $a=b$ we can take the Fourier transform on 
$\bfq$ to get the local potential

\eq
F_1(r)=\int d\bfq e^{i\bfq\cdot\bfr}
e^{-\h a_1q^2}=\left(\frac{2\pi}{a_1}\right)^{\frac{3}{2}}e^{-\h r^2/a_1}.
\qe

In the general case we can write the Fourier transform as
\eq
F(\bfr,\bfr')=\frac{1}{(2\pi)^6}\int d\bfk d\bfk' 
e^{-i\bfk\cdot\bfr}
e^{i\bfk'\cdot\bfr'} 
e^{-\frac{a}{2}(k^2+k'^2)+\bfk\cdot\bfk'b}
\qe
\eq
=\frac{1}{(2\pi)^6}\int d\bfkap d\bfq 
e^{-i\bfkap\cdot(\bfr-\bfr')}
e^{-i\bfq\cdot(\bfr+\bfr')/2} 
e^{-\kappa^2(a-b)}e^{-q^2(a+b)/4}.
\qe
We use the formula
\eq
\int d\bfr e^{i\bfq\cdot\bfr} e^{-\alpha^2 r^2}
=\frac{\pi^{\frac{3}{2}}}{\alpha^3}e^{-\frac{q^2}
{4\alpha^2}}\label{basictransform}
\qe
to do the two Gaussian integrals separately to obtain
\eq F(\bfr,\bfr')=
%\frac{1}{(2\pi)^6}
%\frac{\pi^{\frac{3}{2}}} {(a-b)^{\frac{3}{2}}}
%e^{-\frac{(\bfr-\bfr')^2}{4(a-b)}} 
%\frac{8\pi^{\frac{3}{2}}} {(a+b)^{\frac{3}{2}}}
%e^{-\frac{(\bfr+\bfr')^2}{4(a+b)}}
\frac{1}{(2\pi)^3\sqrt{a^2-b^2}^3}
e^{-\frac{(\bfr-\bfr')^2}{4(a-b)}} 
e^{-\frac{(\bfr+\bfr')^2}{4(a+b)}}
\qe
\eq
%=\frac{1}{(2\pi)^3\sqrt{a^2-b^2}^3}e^{-\frac{a(r^2+r'^2)}
%{2(a^2-b^2)}}e^{\frac{b\bfr\cdot\bfr'}{a^2-b^2}}
=\frac{4\pi}{(2\pi)^3\sqrt{a^2-b^2}^3}e^{-\frac{a(r^2+r'^2)}
{2(a^2-b^2)}}
\sum_{\ell=0}^{\infty} Y_{\ell}^{m^*}(\hat{\bfr})
Y_{\ell}^{m}(\hat{\bfr}')Q_{\ell}\left(
\frac{brr'}{a^2-b^2}\right)
\qe
where
\eq
Q_{\ell}(z)=(-i)^{\ell}j_{\ell}(iz)
\qe
and $j_{\ell}(iz)$ is the spherical Bessel function of imaginary
argument. The local potential can be recovered by taking the limit
$b\rightarrow a$ in the non-local expression.

The wave equation to be solved is
\eq
\nabla^2\psi(\bfr)+\frac{2m_R}{\hbar^2}V_1(r)\psi(\bfr)
+\frac{2m_R}{\hbar^2}\sum_{i=2,3,4}\int d\bfr'V_i(\bfr,\bfr')
\psi(\bfr')=k_0^2\psi(\bfr)
\qe
from which the solution for the wave function and phase shifts can be 
obtained by standard 
matrix techniques.

Because local potentials are relative easy to deal with, a 
commonly used approximation has been to create a local potential 
which is energy dependent although it is inconvenient due to the 
loss of orthogonality of wave functions at different energies. We 
considered such a procedure in this case. Thus we could write
\eq
F(\bfk,\bfk')=e^{-\frac{a-b}{2}(k^2+k'^2)-\frac{b}{2}(\bfk-\bfk')^2}
\longrightarrow e^{-(a-b)k_0^2-\frac{b}{2}(\bfk-\bfk')^2}
\qe
where $k_0$ is the free-space wave number. However, for the standard
parameters given in Ref. \cite{bs} for the fourth diagram, the value of
$b_4$ is negative so one cannot create a local potential for this
term in this way.  We then abandoned this procedure and continued with
the non-local potential for the terms beyond the first. Of course, an
energy-dependent local potential can be generated for the sum of the
four terms after the local/non-local system has been solved.

\begin{table}
$$
\begin{array}{|rrrrrrrrr|}
\hline
&\vline&\alpha\ \ \ \ \ &R\ (N) 
&\beta\ \ \ \ \ &R (K)\ &F\ \ &{\rm 
Born}\ \ \ \ \ &{\rm Potential}\ \ \ \\
{\rm Set}&{\rm Description}\ \vline&{\rm (GeV/c)}\ 
&{\rm (fm)}\ \ \ \ & {\rm (GeV/c)}\ &{\rm (fm)}\ \ \ \ 
& &a_0,a_1\ {\rm (fm)}\ 
&a_0,a_1\ {\rm (fm)}\\
\hline
a&{\rm Ref.\ set\ }\ \vline&0.40\ \ \ &0.85\ \  
\ &0.35\ \ \ &0.49\ \ \ &1.00&-0.12,\ 
-0.35 &-0.10,\ -0.23\\
%\hline
b&{\rm Fit\  }\ \vline&0.68\ \ \ 
&0.50\ \ \ &0.43\ \ \ &0.40\ \ \ &0.98&-0.15,\ 
-0.31&-0.11,\ -0.18\\
%\hline
%c&{\rm Free\ fit}\ \vline&0.46\ \ \ &0.74\ \ \ &0.62\ \ 
%\ &0.28\ \ \ &2.25&-0.15,\ -0.86&-0.11,\ -0.33\\
%\hline
a'&{\rm Mod.\ Ref.}\ \vline&0.40\ \ \ &0.85\ \ \ &0.50\ 
\ \ &0.34\ \ \ &2.06&-0.16,\ -0.78&-0.12,\ 
-0.34\\ 
%\hline
b'&{\rm Mod.\ Fit}\ \vline&0.68\ \ \ &0.50\ \ \ &1.15\ \ \ 
&0.15\ \ \ &4.84&-0.21,\ -1.89&-0.11,\ -0.31\\ 
%\hline
c&{\rm Lattice\ \ }\vline&0.61\ \ \ &0.56\ \ \ 
&0.93\ \ \ &0.18\ \ \ &3.36&-0.18,\ -1.32&-0.11,\ -0.31\\ 
\hline
\end{array}$$
\caption{Parameter values for the single-Gaussian fit to the quark 
density. $R(N)$ is the rms value of the inter-quark distance in the 
nucleon and $R(K)$ is the same quantity for the kaon. Also given is the 
original reference set as well as the 
fitted set of Ref. \protect{\cite{bs}}. Sets $a'$ and $b'$ are the result 
of holding the nucleon size fixed at the unprimed values and adjusting
$\beta$ and $F$ to fit the phase shifts.}
\label{table1}
\end{table}

\section{Results for single Gaussian}

\begin{figure}[htb]
\epsfig{file=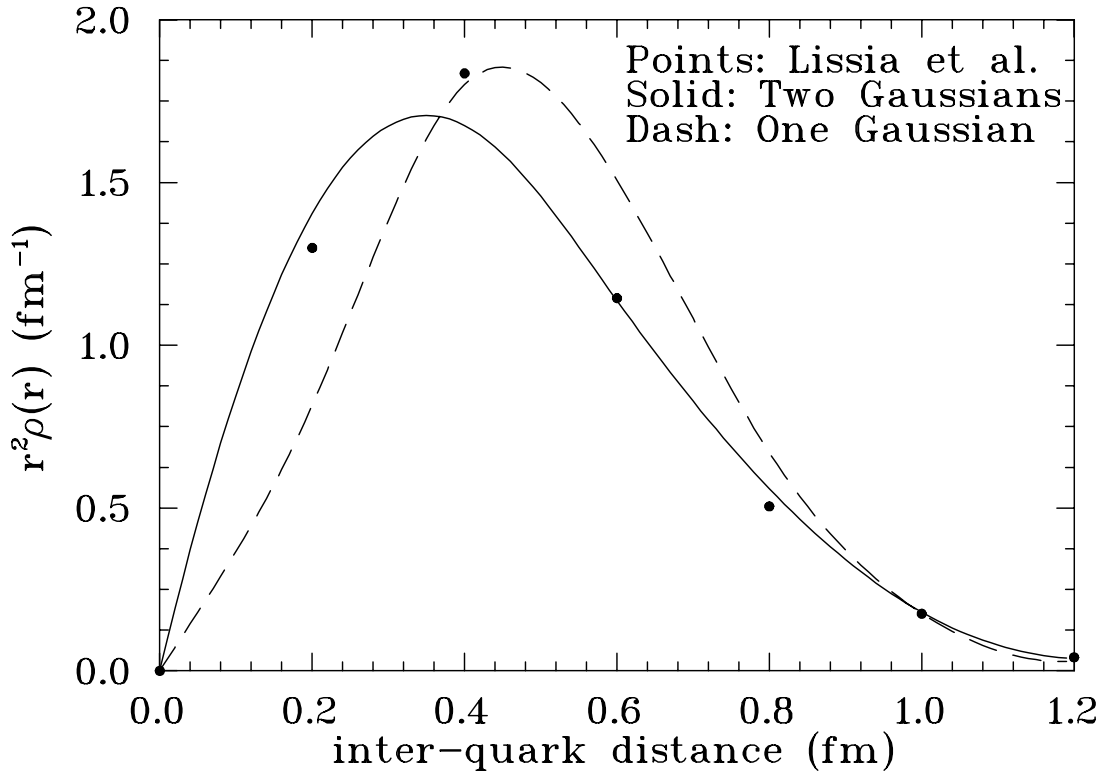,angle=90,height=4.in}
\caption{One and two Gaussian fit to the proton wave function. The points 
are from Lissia \ea\cite{lissia}.} \label{pwf} 
\end{figure}

Many different assumptions have been made about the spacial extent 
of the valence quark distribution of the nucleon so that we need 
to consider a possible range of models.  One approach which should 
be useful is the calculation on the lattice. We will use that of 
Lissia \ea \cite{lissia} although the results of Alexandrou \ea 
\cite{alex} are similar. Figure \ref{pwf} shows (dotted line) the 
result of a fit of single Gaussian to the data from Lissia \ea 
\cite{lissia}.  For this single-Gaussian case we can use the 
equations from Ref. \cite{bs} directly to compute the amplitude. 
Since the light quark mass and the strong coupling constant enter 
only in the combination, $\gamma=\alpha_s/m_s^2$, rather than 
consider changing these quantities independently we calculate 
relative to a standard value of $\gamma_0=0.6/0.33^2=5.51\ {\rm 
(GeV/c^2)^{-2}}$. We define $F=\gamma/\gamma_0$ and keep the ratio 
of light to strange quark fixed at $\rho=0.6$.

\begin{figure}[htb]
\epsfig{file=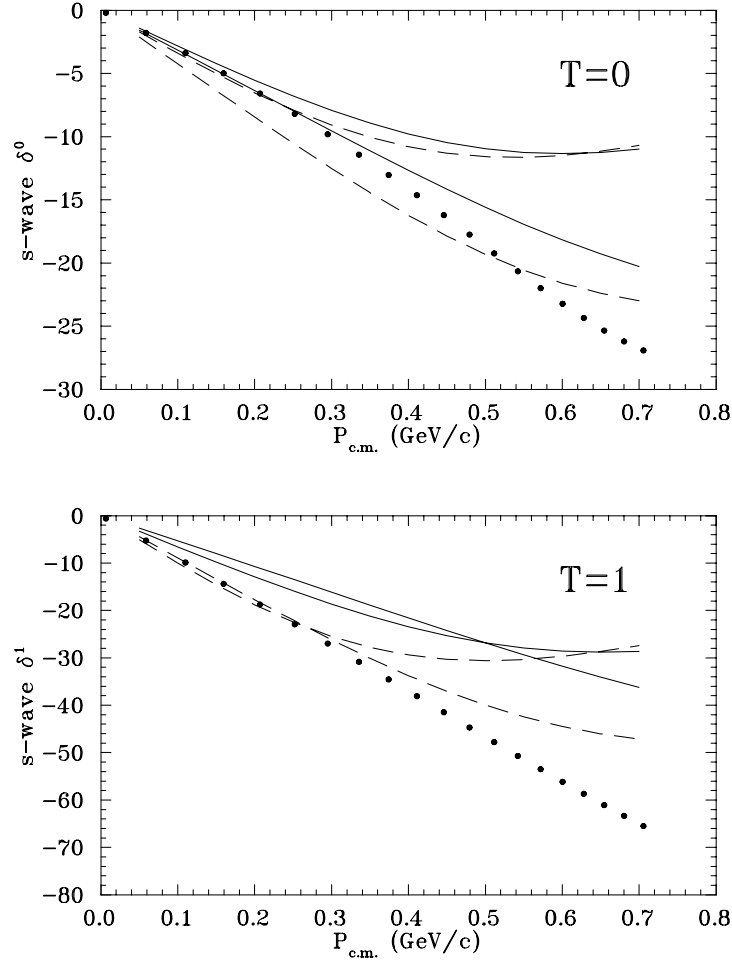,height=5.in}
\caption{Results of Barnes and Swanson with their reference 
and fitted set in Born approximation (dashed lines) and the corresponding 
phase shifts from the solution with the potential (solid lines).
The phase shifts from Ref. \cite{ga} are shown by the solid dots.} 
\label{barnespcmga} \end{figure}

\begin{figure}[htb]
\epsfig{file=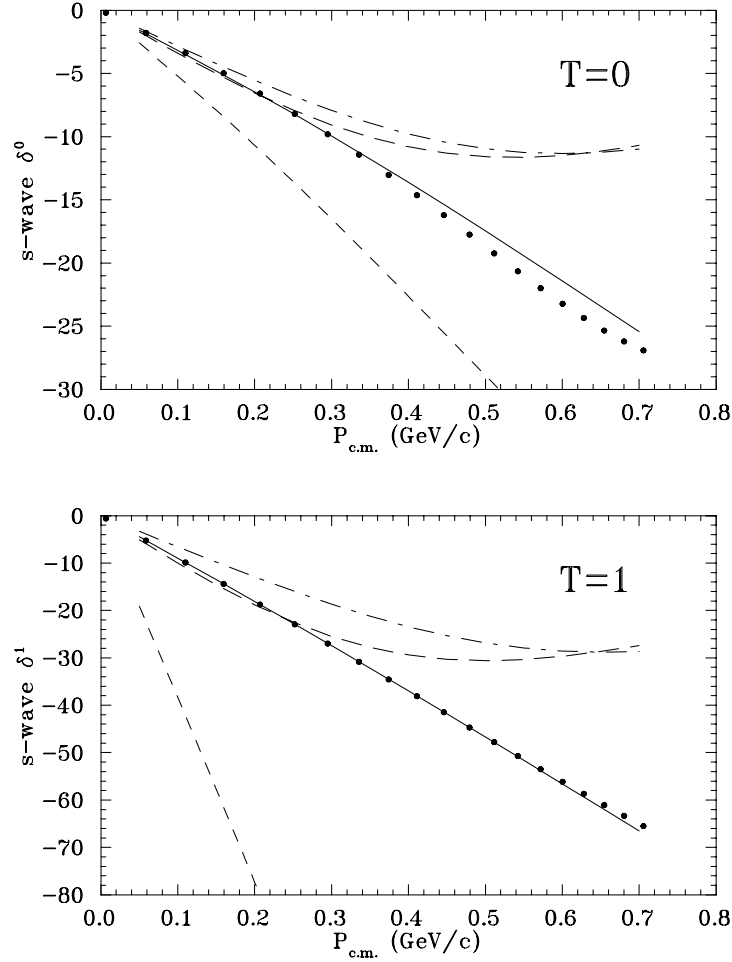,height=5.in}
\caption{S-wave phase shifts using a wave function with a single value of 
$\alpha$ obtained from the fit to Lissia et al. \cite{lissia}. Born 
approximation (short dashed lines) and the corresponding phase shifts from 
the solution with the potential (solid lines). The result of the reference 
calculation and its potential corrected values are shown by the long dash 
and dash dot curves respectively. The phase shifts from Ref. \cite{ga} are 
shown by the solid dots.}
\label{barnespcm} \end{figure}

\begin{figure}[htb]
\epsfig{file=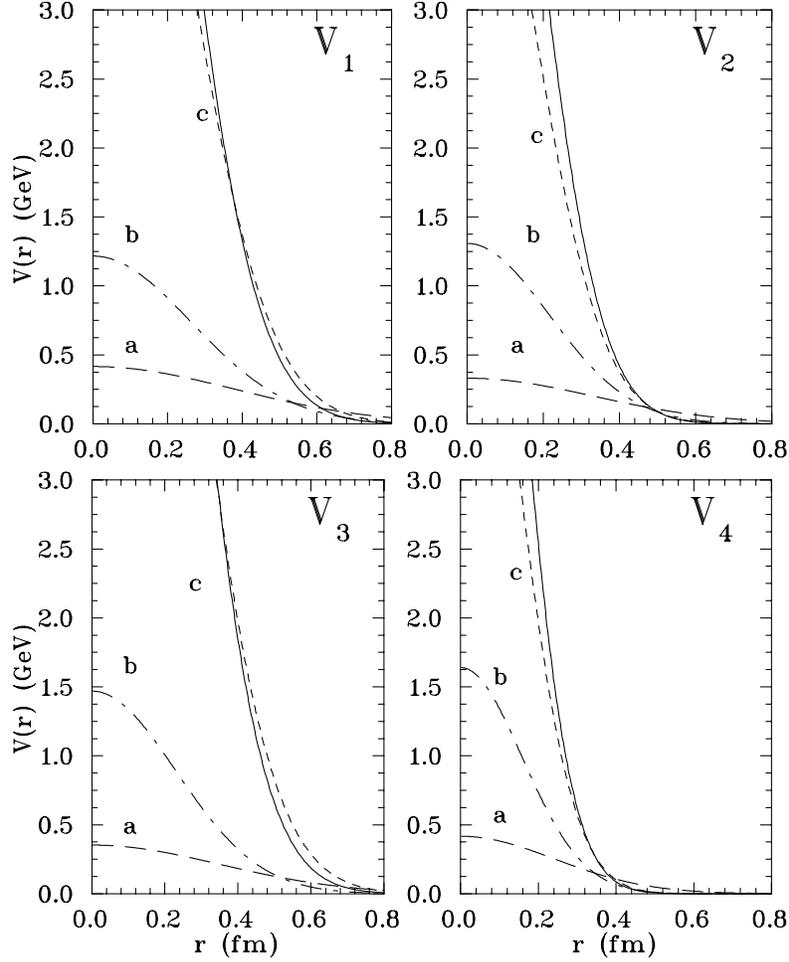,height=5.in}
\caption{Potentials without the isospin factors for K$^+$N scattering for 
three of the parameter sets given in Table \ref{table1}. The upper left 
hand panel shows the local potential corresponding to the first diagram 
and the rest show the corresponding equivalent local potentials at zero 
energy. The short-dash curve with $c$ shows the effect of changing 
$\alpha$ from 0.61 to 0.55 GeV/c.} 
\label{pots} 
\end{figure}

\begin{figure}[htb]
\epsfig{file=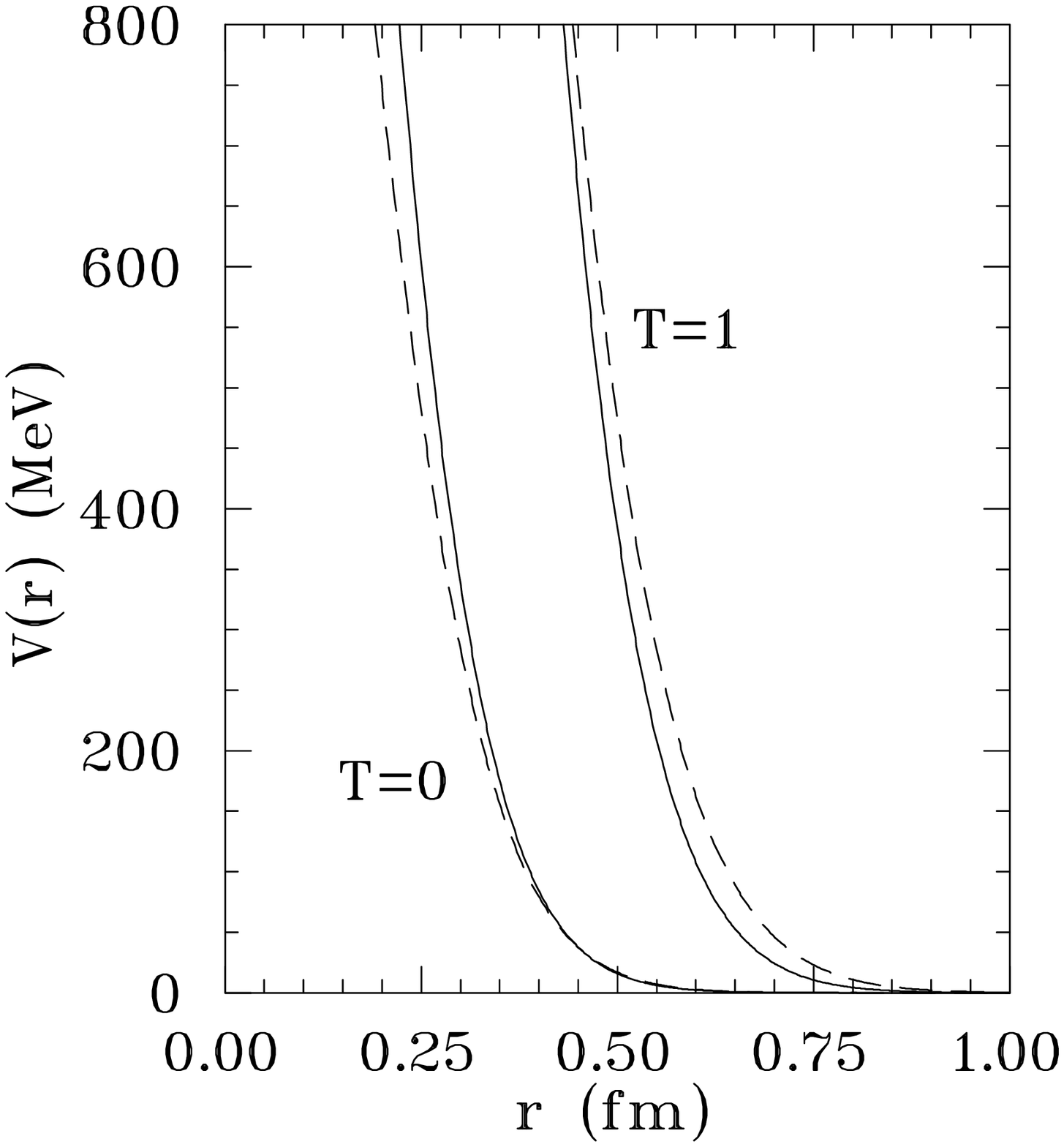,height=4.in}
\caption{The full isospin potentials for a single Gaussian at zero 
energy using the equivalent local forms for the non-local 
potentials calculated with the parameter set $c$ in Table 
\ref{table1} (solid curve) and with $\alpha=0.55$\ GeV/c (dashed 
curve).}
\label{fullpot} 
\end{figure}

Reference \cite{bs} first calculated with a ``reference'' set of 
parameters taken from quark systematics. For this ``reference'' 
set we find that, although the scattering lengths for the Born 
approximation are in agreement with the amplitude 
analysis\cite{ga}, the potential solution is not. The phase shifts 
from this calculation are shown in Fig. \ref{barnespcmga}. 
Although the Born result has the correct ratio of iso-triplet to 
iso-singlet scattering length, the potential result does not. 
Since this ratio is sensitive to the relative hadron sizes we can 
change the kaon size parameter to $\beta=0.5\ {\rm GeV/c}$ to get 
the correct ratio in the potential calculation and increase the 
strength of the potential by a factor of 2.06 to get the right 
magnitude. In this manner a good fit for the potential calculation 
is found (see case $a'$ in Table \ref{table1}).

If we use the ``fitted'' values obtained by Ref. \cite{bs} we find 
case $b$ in Table \ref{table1}. For this value of the nucleon size 
we need to alter the strength of the potential by a factor of 4.84 
and choose $\beta=1.15\ \Gc $ to get the correct values for the 
scattering lengths (Table \ref{table1}).

If we fix the value of $\alpha=0.61\ \Gc$ from the single Gaussian 
fit to Lissia \ea, increase the strength of the potential by a 
factor of 3.36 and take $\beta=0.93\ \Gc$ we are able to fit the 
scattering phase shifts well (case $c$ in Table \ref{table1}). 
Figure \ref{barnespcm} shows the result of this calculation along with
the ``reference'' set of Ref. \cite{bs}.

Figure \ref{pots} show the results for the individual diagram 
potentials without isospin factors from Eq. \ref{isofacs} 
(equivalent local potential at zero energy for all but the first) 
for the ``reference'', ``fit'' and lattice input of case $c$. Also 
shown (short dash curve) is the result of changing $\alpha$ from 
0.61 GeV/c to 0.55 GeV/c ($R_{rms}=0.56$\ fm to $R_{rms}=0.62$\ 
fm).

It is seen that the ranges of the potentials from the second and 
fourth diagrams (the only ones which contribute to the $T=0$ 
amplitude) appear to decrease with increasing nucleon radius 
whereas the range of the potential from the first and third (which 
dominate the isospin one amplitude) increase. One can trace this 
behavior back to the quantities $\eta_i$, $a_i$, and $b_i$. In 
fact, the ``true'' range parameters ($a_i,\ b_i$) always increase 
with increasing $R_{rms}$ but $\eta_2$\ and $\eta_4$ decrease 
rapidly.

 Figure \ref{fullpot} shows the full potentials for the two 
isospins for case $c$. It is seen that the potential is very
near to a hard-core potential for case $c$. This result can no 
doubt be traced to the fact that for both isospins the phase 
shifts are nearly linear as a function of the center-of-mass 
momentum as is the case for a hard-sphere interaction. Also shown 
(short dash curve) is the result for $\alpha=0.55$\ GeV/c. As 
expected from the remarks above, the isospin zero potential has 
the opposite variation with nucleon size as the isospin one 
potential.

These results show that the Born approximation (especially for $T=1$)
is not accurate. The use of a potential (or other unitarizing agent) is 
essential. The range of the potential (related to the size of the 
hadronic systems) is important. For a more extended potential the Born
approximation is more nearly correct.
 
\section{Two Gaussian Form of the Wave Function}

With the extention of the formulas of Ref. \cite{bs} given in the 
next section one could carry out the calculation of a potential 
using a sum of any number of Gaussian functions for the quark wave 
function of the nucleon. A similar program was carried out for 
meson-meson scattering by Hilbert \ea\cite{hilbert}. Of course, 
such a sum would introduce a large number of parameters which have 
to be determined. Since we intend to fit them to lattice data, and 
those calculations are of a limited accuracy at present, we use a 
sum of only two such functions.  The wave functions below are 
expressed in Jacobian coordinates as discussed in Appendix A.

We take the sum of two Gaussian wave functions, each normalized 
independently. With
\eq
\psi(p,q)=N_a\phi_a(p,q)+N_b\phi_b(p,q)
\qe
we have for the normalization integral 
\eq
\int d\bfp d\bfq \psi^2(p,q)=27(N_a^2+N_b^2)+2N_aN_b\int d\bfp d\bfq
\phi_a(p,q)\phi_b(p,q)=1.
\qe
Since 
\eq
\int d\bfp d\bfq \phi_a(p,q)\phi_b(p,q)=\frac{3^{\frac{3}{2}}(4\pi)^2}
{\pi^3\alpha_a^3\alpha_b^3}\int p^2dpq^2dq 
e^{-\frac{p^2}{4}\left(\frac{1}{\alpha_a^2}+\frac{1}{\alpha_b^2}\right)}
e^{-\frac{q^2}{3}\left(\frac{1}{\alpha_a^2}+\frac{1}{\alpha_b^2}\right)}
=27\frac{8\alpha_a^3\alpha_b^3}{(\alpha_a^2+\alpha_b^2)^3},
\qe
the normalization condition becomes
\eq
N_a^2+\frac{16N_1N_2\alpha_a^3\alpha_b^3}{(\alpha_a^2+\alpha_b^2)^3}
+N_b^2=1.
\qe
If we take $N_b=X N_a$ then
\eq
N_a^2\left[1+\frac{16X\alpha_a^3\alpha_b^3}{(\alpha_a^2+\alpha_b^2)^3}
+X^2\right]=1.
\qe
We are free to take $X$, $\alpha_a$ and $\alpha_b$ as we wish and, as
long as this condition is satisfied, the wave function will be normalized.

\section{Summary of formulas}

In order to carry out the calculation of the potential (or the amplitude) 
for a sum of Gaussian forms for the wave function we must evaluate the 
integrals given in Ref. \cite{bs} (Eqs. 41, 42, 43 and 44 which are our 
equations \ref{dia1}, \ref{dia2}, \ref{dia3} and \ref{dia4}) for the case 
where kaons A and C have different governing Gaussian parameters ($\beta$) 
and nucleons B and D also have different parameters ($\alpha$). The 
evaluation of the integrals is done in appendix B and here we give a 
summary of the results. Auxiliary definitions are given at the end.

\newpage

\eq
a_1=\frac{1}{2\tbe^2(1+\rho)^2}+\frac{1}{3\tal^2};\ \  b_1=a_1;\ \ 
\eta_1=\frac{\alpha_1^3\alpha_2^3\beta_1^{\frac{3}{2}}\beta_2^{\frac{3}{2}}}
{\tal^6\tbe^3}
\qe
\eq
a_2=\frac{1}{2(1+\rho)^2\tbe^2}+\frac{1}{12\tal^2}+
\frac{C_2^2+C'^2_2}{4(\alpha_1^2+2\beta_{12}^2+\frac{1}{3}
\alpha_{12}^2)}
\qe
$$
b_2=\frac{1}{2(1+\rho)^2\tbe^2}+\frac{1}{12\tal^2}-
\frac{C_2C'_2}{2(\alpha_1^2+2\beta_{12}^2+\frac{1}{3}\alpha_{12}^2)}
$$
\eq C_2=
-\frac{[(2+\gamma^{\alpha}_1)(1+\rho)-6\gamma^{\beta}_2]}{3(1+\rho)}
;\ \ 
C'_2=\frac{[-\gamma^{\alpha}_1(1+\rho)+6(\gamma^{\beta}_2+\rho)]}
{3(1+\rho)}
\qe

\eq
\eta_2=\frac{2^{\frac{3}{2}}\alpha_1^3\alpha_2^3\beta_1^{\frac{3}{2}}
\beta_2^{\frac{3}{2}}}{\tbe^3\tal^3(\alpha_1^2+2\beta_{12}^2+\frac{1}{3}
\alpha_{12}^2)^{\frac{3}{2}}}
\qe

\eq
a_3=\frac{1}{3\tal^2}+\frac{3(C_3^2+C'^2_3)}{8(2\alpha_{12}^2+3\beta_1^2)}
;\ \ 
b_3=\frac{1}{3\tal^2}-\frac{3C_3C'_3}{4(2\alpha_{12}^2+3\beta_1^2)}
\qe
\eq C_3= 
\frac{2[2\gamma^{\alpha}_2+(2\gamma^{\alpha}_2-3)\rho]}{3(1+\rho)};\ \ 
C'_3= -\frac{2}{3}(2\gamma^{\alpha}_2+1)
\qe

\eq \eta_3=\rho\frac{
6^{\frac{3}{2}}\alpha_1^3\alpha_2^3\beta_1^{\frac{3}{2}}
\beta_2^{\frac{3}{2}}}{\tal^6(2\alpha_{12}^2
+3\beta_1^2)^{\frac{3}{2}}}
\qe
\eq a_4=\frac{(1-\rho)^2}{4(\tal+2\tbe)(1+\rho)^2}+\frac{1}{12\tal^2}
+\frac{C_4^2+C'^2_4}{4\left(\frac{v^2w^2}{v^2+w^2}+\frac{1}{3}
\alpha_{12}^2\right)}
\qe
\eq b_4=\frac{(1-\rho)^2}{4(\tal+2\tbe)(1+\rho)^2}+\frac{1}{12\tal^2}
-\frac{C_4C'_4}{2\left(\frac{v^2w^2}{v^2+w^2}+\frac{1}{3}
\alpha_{12}^2\right)}
\qe   
\eq C_4=-\frac{2w^2-v^2+\frac{6a^2\rho}{1+\rho}+(v^2+w^2)
\gamma^{\alpha}_1}{3(v^2+w^2)};\ \ 
C'_4=-\frac{\frac{6w^2\rho}{1+\rho}+3v^2-(v^2+w^2)\gamma^{\alpha}_1}
{3(v^2+w^2)}
\qe
\eq 
\eta_4=\rho\alpha_1^3\alpha_2^3\beta_1^{\frac{3}{2}}\beta_2^{\frac{3}{2}}
\left(\frac{8}{(v^2+w^2)\tal^2}\right)^{\frac{3}{2}}
\frac{1}{\left(\frac{v^2w^2}{v^2+w^2}+\frac{1}{3}\alpha_{12}^2
\right)^\frac{3}{2}}
\qe

\eq
v^2=\alpha_1^2+2\beta_2^2;\ \ w^2=\alpha_2^2+2\beta_1^2
\qe
\eq
\gamma^{\beta}_i=\frac{\beta_i^2}{\beta_1^2+\beta_2^2};\ \ \ 
\gamma^{\alpha}_i=\frac{\alpha_i^2}{\alpha_1^2+\alpha_2^2};\ \ \ 
\qe

\eq
\tilde{\alpha}^2=\h(\alpha_1^2+\alpha_2^2);\ \ 
\tilde{\beta}^2=\h(\beta_1^2+\beta_2^2);\ \ 
\alpha_{12}^2=\frac{\alpha_1^2\alpha_2^2}{\alpha_1^2+\alpha_2^2}
\ \ \beta_{12}^2=\frac{\beta_1^2\beta_2^2}{\beta_1^2+\beta_2^2}
\qe

\begin{figure}[htb]
\epsfig{file=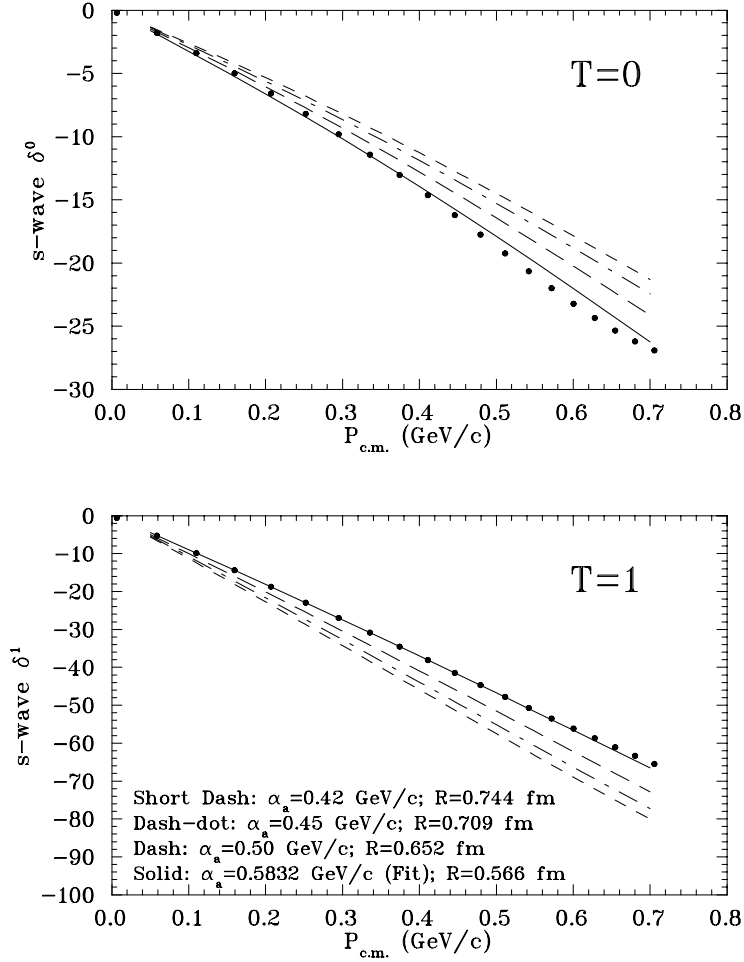,height=5.in}
\caption{S-wave phase shifts with $\alpha_a$, $\alpha_b$ and $x$ chosen 
from Lissia et al. \cite{lissia} and the overall factor on the potential 
chosen to fit the $T$=1 phase shift. The dashed curve shows the 
result if the value of $\alpha_a$ is altered from its fitted value 
corresponding to the changes in density shown in Fig. 
\ref{rhovary}. The phase shifts from Ref. \cite{ga} are shown by 
the solid dots.}
\label{bs2pcmvary} \end{figure}

\begin{figure}[htb]
\epsfig{file=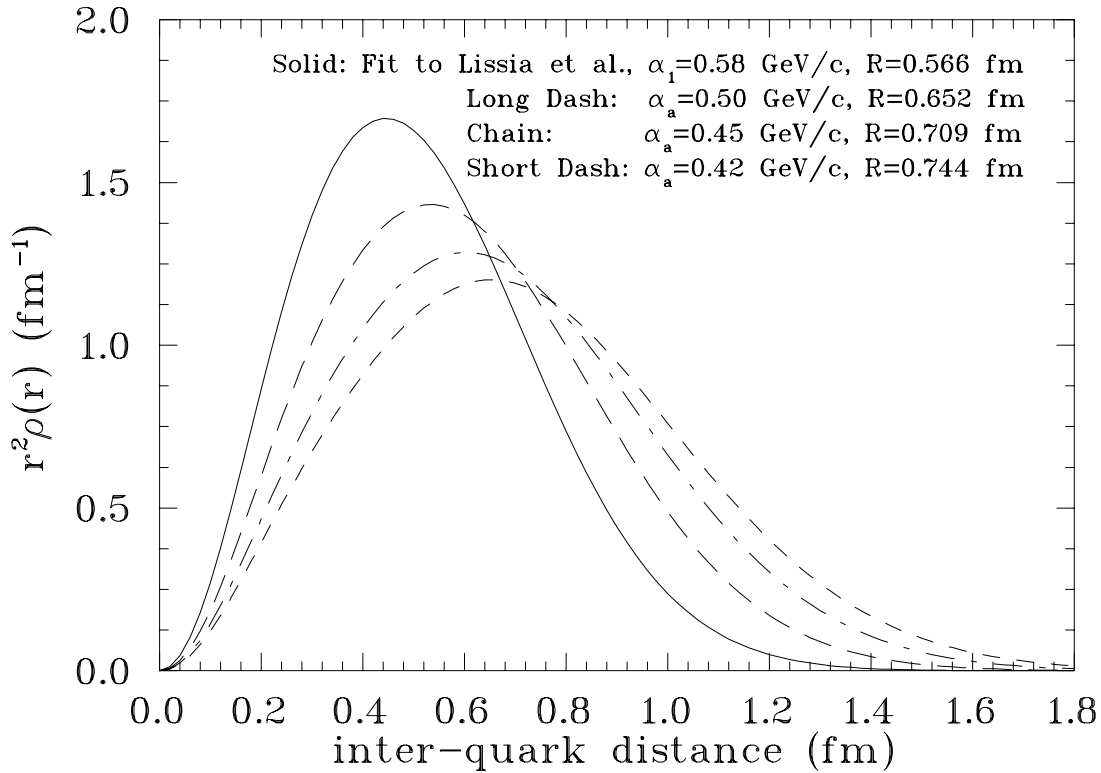,angle=90,height=4.in}
\caption{The solid curve gives the result for $r^2\rho(r)$ with 
$\alpha_a$, $\alpha_b$ and $X$ chosen from Lissia et al. \cite{lissia}.  
The curves show the densities that result for various values of 
$\alpha_a$ with $\alpha_b=1.32$\ GeV/c and $X$ fixed at 0.1. } 
\label{rhovary} \end{figure}

\section{Results for a double Gaussian Wave Function}

Figure \ref{pwf} shows a fit (solid line) of the two-Gaussian type 
to the lattice calculation of Lissia \ea \cite{lissia}.  The 
values of the double Gaussian parameters are $\alpha_a=0.58\ \Gc$, 
$\alpha_b=1.32\ \Gc$ and $X=0.1$. This selection of parameters 
gives a reasonable representation of the s-wave phase shift (see 
Fig. \ref{bs2pcmvary}) but a factor of 4.1 has been multiplied by 
the potential to obtain the fit.

The fit with the double Gaussian is shown in Fig. \ref{bs2pcmvary} 
as the solid line. Also shown are a series of variations of 
$\alpha_1$ (holding $\alpha_2$ and X fixed) used to vary the 
radius of the system. The effect of this variation on the density 
is shown in Fig. \ref{rhovary}.  This variation would be supposed 
to come from the partial deconfinement of the quarks in a nucleus. 
As expected from the remarks on the behavior of the potentials as 
a function of nuclear size for the single Gaussian case, the $T=0$ 
phase shift decreases with increasing nucleon size.

Although the ratio of total cross section of nuclei to that of 
deuterium measured\cite{krauss,weiss} clearly shows a deviation 
from the predicted ratio with free kaon-nucleon interactions, the 
energy dependence is not in agreement with the simple change in 
radius estimated in Ref. \cite{siegel} being larger at higher 
incident momenta. Since the model used to calculate the ratio is 
rather simple, one should consider the possibility that a more 
realistic model could give the energy dependence necessary. For 
this reason we compare the change in cross section that results 
from a given change in nucleon radius for two different energies.  
We also make the same type of comparison for changes in 
interaction strength since that possibility has been suggested as 
an alternative to an increase in nucleon size\cite{brown}.

In a series of papers, Caillon and Labarsouque\cite{caillon} 
studied this problem with different possibilities for the 
interaction of the kaon with the nucleus. They are able to 
partially resolve this problem but have difficulty predicting the 
correct dependence on atomic number.

In Tables \ref{0.35}, \ref{0.5}, \ref{0.35l} and \ref{0.5l} is shown
the effect on the elastic and total cross sections of the variation in
radius and strength. For each isospin the amplitude 
\eq
f_T=\frac{e^{2i\delta_T}-1}{2i}
\qe
is calculated and the isospin weighting is done. Since at finite 
energies a linear expansion of the exponential does not provide a 
good approximation and since the variations in the two isospin 
phase shifts with radius and strength go in opposite directions 
there is not a simple linear proportionality.

From Table \ref{0.35} we see that at a beam momentum of about 0.6 
GeV/c a change in the radius of 15\% corresponds to a change in 
the differential cross section of 14\% and a change in the total 
cross section of 18\%. A change in the radius of 25\% gives 
corresponding changes of 25\% and 32\%. From Table \ref{0.5} we 
see that at a higher beam momentum the same fractional changes in 
radii lead to smaller fractional changes in the cross sections. 
The trend is opposite to what would be needed to give a better fit 
to the data\cite{krauss,weiss}.

\begin{table}
$$
\begin{array}{|cccccccccc|}
\hline
{\rm R (fm)}&{\rm R/R^0}&\delta_0
&\delta_1 &f_0&f_1&\frac{3}{4}f_1+\frac{1}{4}f_0&k^2\sigma
&\sigma/\sigma^0&\sigma_T/\sigma_T^0\\
\hline
0.566&1.00&-0.210&-0.562&-0.204+i0.043&-0.451+i0.284&-0.389+i0.224
&0.202&1.00&1.00\\
0.652&1.15&-0.192&-0.623&-0.188+i0.037&-0.474+i0.340&-0.402+i0.264
&0.232&1.15&1.18\\
0.709&1.25&-0.178&-0.669&-0.175+i0.032&-0.486+i0.384&-0.409+i0.296
&0.255&1.26&1.32\\
0.744&1.31&-0.166&-0.700&-0.166+i0.029&-0.493+i0.415&-0.411+i0.318
&0.270&1.34&1.41\\
\hline
\end{array}
$$
\caption{Dependence of the s-wave amplitude on the quark radius at 
P$_{c.m.}=0.35$ GeV/c (P$_{Lab}$=0.599 GeV/c). R is the rms value of
the inter-quark distance in the nucleon. The quantity $\sigma$ is 
the (isoscalar) differential cross section which would come from the 
s-wave alone and the quantity $\sigma_T$ is the (isoscalar) total cross
section from the s-wave alone. The superscript zero refers 
to the first line in the table i.e. the result obtained from the fit to 
the data in free space.}
\label{0.35}\end{table}
\begin{table}
$$
\begin{array}{|cccccccccc|}
\hline
{\rm R (fm)}&R/R^0&\delta_0
&\delta_1 &f_0&f_1&\frac{3}{4}f_1+\frac{1}{4}f_0&k^2\sigma
&\sigma/\sigma^0&\sigma_T/\sigma_T^0\\
\hline
0.566&1.00&-0.312&-0.816&-0.292+i0.094&-0.499+i0.531&-0.447+i0.422
&0.378&1.00&1.00\\
0.652&1.15&-0.287&-0.900&-0.271+i0.080&-0.487+i0.614&-0.433+i0.480
&0.418&1.11&1.14\\
0.709&1.25&-0.267&-0.963&-0.254+i0.069&-0.469+i0.673&-0.415+i0.522
&0.445&1.18&1.24\\
0.744&1.31&-0.253&-1.004&-0.242+i0.063&-0.453+i0.711&-0.400+i0.549
&0.462&1.22&1.30\\
\hline
\end{array}
$$
\caption{Dependence of the s-wave amplitude on the radius at 
P$_{c.m.}=0.5$ GeV/c (P$_{Lab}$=0.941 GeV/c). R is the rms value of the
inter-quark distance in the nucleon. The superscript zero refers 
to the first line in the table i.e. the result obtained from the fit to 
the data in free space.}
\label{0.5}\end{table}

\begin{table}
$$
\begin{array}{|cccccccccc|}
\hline
F&F/F^0&\delta_0
&\delta_1 &f_0&f_1&\frac{3}{4}f_1+\frac{1}{4}f_0&\sigma
&\sigma/\sigma^0&\sigma_T/\sigma_T^0\\
\hline
4.1&1.00&-0.210&-0.562&-0.204+i0.043&-0.451+i0.284&-0.389+i0.224
&0.202&1.00&1.00\\
\hline
5.0&1.22&-0.230&-0.591&-0.222+i0.052&-0.463+i0.311&-0.403+i0.246
&0.223&1.10&1.10\\
\hline
6.0&1.46&-0.248&-0.618&-0.238+i0.060&-0.472+i0.336&-0.414+i0.267
&0.242&1.20&1.19\\
\hline
7.0&1.71&-0.263&-0.640&-0.251+i0.068&-0.479+i0.357&-0.422+i0.284
&0.259&1.28&1.27\\
\hline
8.0&1.95&-0.276&-0.659&-0.262+i0.074&-0.484+i0.375&-0.429+i0.300
&0.273&1.35&1.34\\
%\hline
%9.0&2.20&-0.287&-0.675&-0.271+i0.080&-0.488+i0.391&-0.434+i0.313
%&0.286&1.41&1.40\\
\hline
\end{array}
$$
\caption{Dependence of the s-wave amplitude on the multiplying 
factor at P$_{c.m.}=0.35$ GeV/c (P$_{Lab}$=0.599 GeV/c). The 
superscript zero refers to the first line in the table i.e. the 
result obtained from the fit to the data in free space.}
\label{0.35l}\end{table}

\begin{table}
$$
\begin{array}{|cccccccccc|}
\hline
F&F/F^0&\delta_0
&\delta_1 &f_0&f_1&\frac{3}{4}f_1+\frac{1}{4}f_0&k^2\sigma
&\sigma/\sigma^0&\sigma_T/\sigma_T^0\\
\hline
4.1&1.00&-0.312&-0.816&-0.292+i0.094&-0.499+i0.531&-0.447+i0.422
&0.378&1.00&1.00\\
\hline
5.0&1.22&-0.341&-0.858&-0.315+i0.112&-0.495+i0.572&-0.450+i0.457
&0.411&1.08&1.08\\
\hline
6.0&1.46&-0.367&-0.896&-0.335+i0.129&-0.488+i0.610&-0.450+i0.489
&0.442&1.17&1.16\\
\hline
7.0&1.71&-0.388&-0.927&-0.350+i0.143&-0.480+i0.640&-0.447+i0.516
&0.466&1.23&1.22\\
\hline
8.0&1.95&-0.406&-0.955&-0.363+i0.156&-0.472+i0.666&-0.444+i0.538
&0.487&1.29&1.27\\
%\hline
%9.0&2.20&-0.421&-0.978&-0.373+i0.167&-0.463+i0.688&-0.441+i0.558
%&0.505&1.34&1.32\\
\hline
\end{array}
$$
\caption{Dependence of the s-wave amplitude on the multiplying 
factor at P$_{c.m.}=0.50$ GeV/c (P$_{Lab}$=0.941 GeV/c). The 
superscript zero refers to the first line in the table i.e. the 
result obtained from the fit to the data in free space.}
\label{0.5l}\end{table}

Tables \ref{0.35l} and \ref{0.5l} show a similar effect for the 
variation of the cross sections as a function of the change in the 
multiplying factor. It is seen that a significant change (in 
percentage) of the multiplying factor is needed to produce a 
modest change in the total cross section ratio.

\begin{figure}[htb]
\epsfig{file=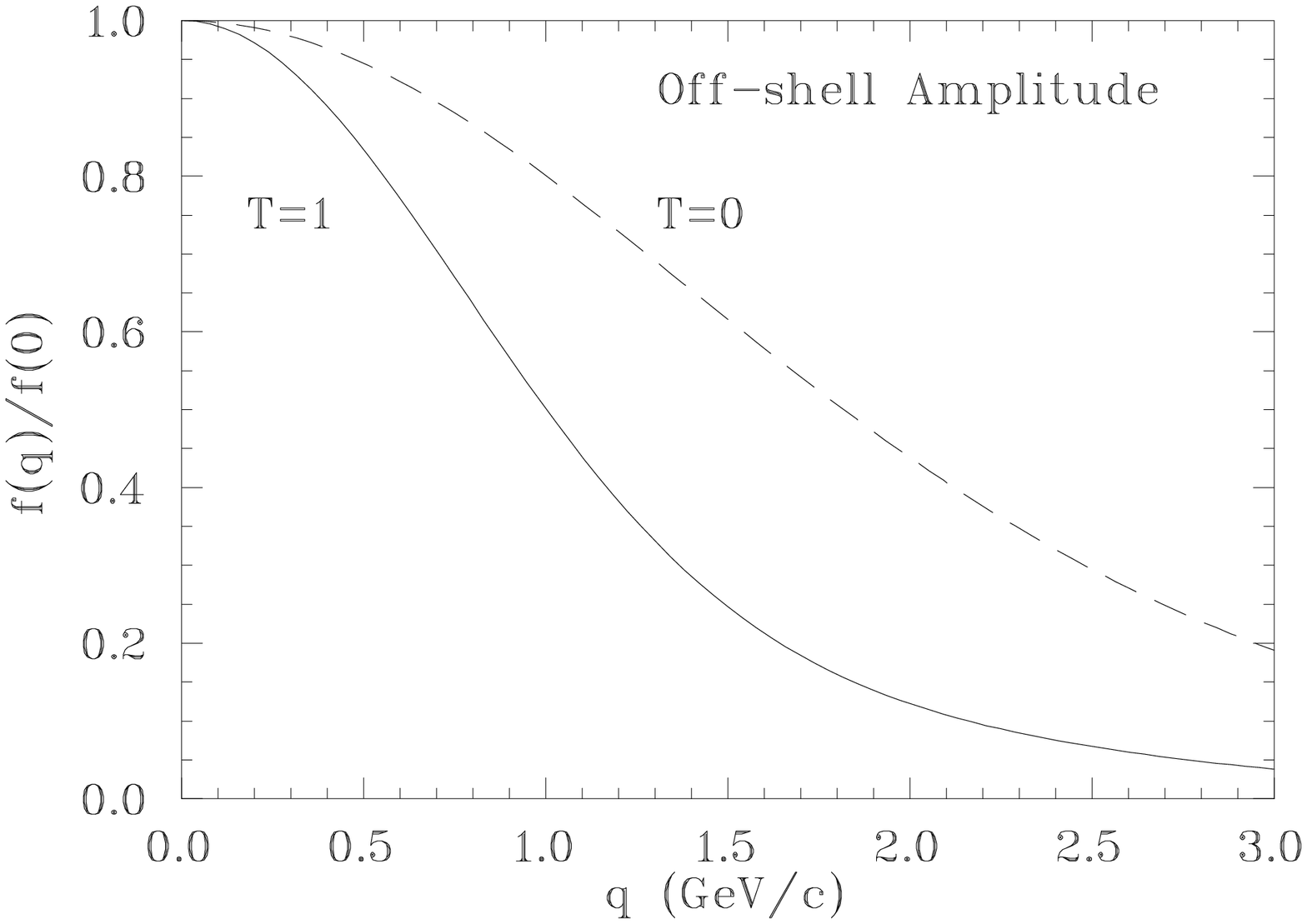,angle=90,height=4.in}
\caption{Ratio of the off-shell to on-shell amplitude at zero energy
for $T$=1 (solid) and $T$=0 (dashed). The function falls to 1/2 
for $T$=1 at about 1 GeV/c. } 
\label{os} \end{figure}

\section{Off-shell Amplitude}

From the full solution for the wave function one can calculate the 
scattering amplitude for off-shell momenta. We have done so at 
(essentially) zero energy. Since the value of the off-shell 
amplitude at zero momenta will be equal to the scattering length, 
dividing this amplitude by the scattering length one obtains a 
function which is unity at $q=0$. Because the wave function is 
constant at zero energy the result is equivalent to the Fourier 
transform of the sum of the four potentials.

Figure \ref{os} shows the result of such a calculation. We see 
that the half value of the $T$=1 function occurs at around 1 
GeV/c.  $T$=1 dominates nuclear scattering because of the average 
weight in the isoscalar average and the relative smallness of the 
$T$=0 scattering length so its range is perhaps the appropriate 
one to use.

\section{Discussion and Conclusions}

We have calculated the kaon-nucleon scattering amplitude as a 
function of the nucleon size and interaction strength in the 
one-gluon-exchange approximation. Two interesting features are 
observed:

1) The isospin one and isospin zero amplitudes do not behave in 
   the same manner under a change in nucleon size. The isospin 
   zero amplitude actually decreases with increasing nucleon size. 
   Although the potential used is not unique, the fact that the 
   behavior is due to the sharp decrease of $\eta_2$\ and 
   $\eta_4$\ with increasing nucleon size leads us to believe that 
   the behavior is more general.

2) A significant renormalization of the potential is required to 
   fit the experimental amplitudes. The factor of 4.1 on the 
   strength (relative to a fixed value of $\gamma=\alpha_s/m_q^2$) 
   of the interaction necessary to bring the quark-gluon exchange 
   into agreement with the s-wave phase shifts is considerable.  
   We have found that the factor needed depends on the extent of 
   the quark distribution of the nucleon and kaon. If this result 
   is confirmed then we must expect additional quark diagrams to 
   play an important role or, alternatively, the exchange of color 
   singlets (for example two-pion exchange with excitation of the 
   K$^*$) may be important. The very small range of the kaon 
   distribution is also disturbing. Because the potential which 
results from the fit is very near to that of a hard-sphere, one 
might think that Pauli blocking might play an important 
role\cite{bender}.

We thank Ted Barnes and Eric Swanson for valuable correspondence
concerning this work.

This work was supported by the National Science Foundation under
contract PHY-0099729.

\appendix

\section{Jacoby Coordinates}

The integrals needed for the calculation are more easily done if the 
momentum variables are expressed in terms of Jacoby coordinates.
Barnes and Swanson\cite{bs} use a nucleon wave function 
represented as

\eq  \phi(\bfp_1,\bfp_2,\bfp_3)=\frac{3^{\frac{3} {4}}} 
{\pi^{\frac{3}{2}}\alpha^3}
e^{-\frac{\bfp_1^2+\bfp_2^2+\bfp_3^2-\bfp_1\cdot\bfp_2
-\bfp_2\cdot\bfp_3-\bfp_3\cdot\bfp_1}{3\alpha^2}}.
\qe

Expressed in terms of coordinates
\eq
\bfp\equiv \bfp_1-\bfp_2;\ \ {\bf Q}\equiv 
\frac{\bfp_1+\bfp_2}{2};\ \ \bfq\equiv\bfp_3-{\bf Q}
\label{jacoby}\qe
the wave function becomes
\eq \phi(\bfp,\bfq)
%\propto
%e^{-\frac{\frac{3}{4}p^2+q^2}{3\alpha^2}}\propto 
=\frac{3^{\frac{3} {4}}} {\pi^{\frac{3}{2}}
\alpha^3}e^{-\frac{p^2}{4\alpha^2}}
e^{-\frac{q^2}{3\alpha^2}}.\label{a3}
\qe

We can transform these wave functions into coordinate space.
In the Fourier transform we have
\eq
\bfp_1\cdot\bfr_1+\bfp_2\cdot\bfr_2+\bfp_3\cdot\bfr_3
={\bf Q}\cdot\bfR+\h \bfp\cdot\bfr+\bfq\cdot\bfr_3
\qe
where
\eq
\bfR=\bfr_1+\bfr_2+\bfr_3;\ \ \bfr=\bfr_1-\bfr_2.
\qe
Thus the dependence in r-space is
\eq
\phi(r,r_3)\propto e^{-\frac{\alpha^2 
r^2}{4}}e^{-\frac{3\alpha^2 
r_3^2}{4}}
\qe
giving
\eq
<r^2>\equiv R^2(N)=3/\alpha^2
\qe
for the expectation value of the square of the inter-quark 
separation in the density.

The normalization in this form follows from the normalization
given\cite{bs}
\eq
\int d\bfp_1d\bfp_2d\bfp_3\phi^2(\bfp_1,\bfp_2,\bfp_3)
\delta(\bfP_{c.m.}-\bfp_1-\bfp_2-\bfp_3)
\delta(\bfP'_{c.m.}-\bfp_1-\bfp_2-\bfp_3)
\qe
$$
=\int d\bfp d\bfQ d\bfq\phi^2(\bfp,\bfQ)
\delta(\bfP_{c.m.}-3\bfQ-\bfq)
\delta(\bfP'_{c.m.}-3\bfQ-\bfq)
=\frac{\delta(\bfP_{c.m.}-\bfP'_{c.m.})}{27}\int d\bfp d\bfq
\phi^2(p,q).
$$
The integral over $\bfp$ and $\bfq$ should give 27.
\eq
\int d\bfp d\bfq \phi^2(p,q)=
(4\pi)^2\frac{3^{\frac{3}{2}}}{\pi^3\alpha^6}
\int p^2dpe^{-\frac{p^2}{2\alpha^2}}\int q^2dq 
e^{-\frac{2q^2}{3\alpha^2}}=27.
\qe

The kaon wave function is written in terms of the relative 
momentum as
\eq 
\phi_{kaon}(p)=\frac{1}{\pi^{\frac{3}{4}}\beta^{\frac{3}{2}}
}e^{-\frac{p^2}{8\beta^2}}.
\qe

\section{Derivation of the Amplitude Expressions}

We calculate the expression of Ref. \cite{bs} for the case where the 
initial and final Gaussian parameters are not the same in order to
enable a representation of the wave functions as a sum of Gaussian 
functions.

The strategy is the following:

a) Rewrite the nucleon wave functions in Jacoby coordinates
in the form of Eq. \ref{a3}.

b) Transform to r-space to get the proper combinations
of the momenta. 

c) Perform the momentum integrals to find 3 $\delta$-functions in
$\bfr$

d) Carry out the remaining 3 Fourier transforms.

One has two quantities to follow as the integrals are done:

1) The imaginary part of the exponent and 

2) The real parts of the exponent which represent the Gaussian 
   functions and, when the integral is completed, give the 
   dependence of the amplitude in the form of $a_i$,\ and $b_i$.

In this process the factors in front of the exponentials are 
ignored. They are obtained most easily by setting $\bfk=\bfk'=0$ 
and performing the momentum space integrals directly. The results 
for these normalizations ($\eta_i$) are given in the summary.

The basic Fourier transform used is given by Eq. \ref{basictransform}.
A useful transformation is 
\eq
\int d\bfr e^{i\bfu\cdot\bfr} e^{-z_1^2r^2}e^{-z_2^2(\bfr+\bfr')^2}
=e^{-i\xi\bfu\cdot\bfr'}e^{-z_{12}^2r'^2}
\int d\bft e^{i\bfu\cdot\bft} e^{-2\tilde{z}^2t^2}
\label{b1}\qe
obtained with the change of variables
\eq
\bft=\bfr+\xi\bfr',\ \ {\rm with}\ \ \xi=\frac{z_2^2}{z_1^2+z_2^2}
\qe
and
\eq
z_{12}^2=\frac{z_1^2z_2^2}{z_1^2+z_2^2};\ \ \tilde{z}^2=\h(z_1^2+z_2^2).
\qe
If $z_1=z_2=z$ then $\xi=\h$ (this is the case for the same 
initial and final Gaussian wave functions) the formula simplifies 
to:
\eq
e^{-\h\bfu\cdot\bfr'}e^{-\h z^2r'^2}\int d\bft e^{i\bfu\cdot\bft}
e^{-2z^2t^2}.
\qe

\subsection{Diagram 1}

In this case the integral to be calculated is

$$
\int d\bfa d\bfb_1 d\bfb_2 \phi_A\left(2\bfa-\frac{2\rho\bfk}{1+\rho}
\right)\phi^*_C\left(2\bfa+\frac{2\bfk'}{1+\rho}-2\bfk\right)
\phi_B\left(\bfb_1,\bfb_2,-\bfk-\bfb_1-\bfb_2\right)
$$
\eq \times
\phi^*_D(\bfb_1+\bfk-\bfk',\bfb_2,-\bfk-\bfb_1-\bfb_2).
\label{dia1}\qe
Here $\phi_A$ (with Gaussian parameter $\beta_1$)
and $\phi_C$ (with Gaussian parameter $\beta_2$) represent the two kaon 
wave functions and $\phi_B$ (with Gaussian parameter $\alpha_1$) and
$\phi_D$ (with Gaussian parameter $\alpha_2$) represent the two nucleon
wave functions.

One sees immediately that the
kaon wave function integral on $\bfa$ factors off so that 
we could do the kaon integral directly in momentum space
but to be uniform in procedure we transform to r-space so we have
\eq
\int d\bfa e^{-\frac{(\bfa-\frac{\rho \bfk}{1+\rho})^2}{2\beta_1^2}}
e^{-\frac{(\bfa+\frac{\bfk'}{1+\rho}-\bfk)^2}{2\beta_2^2}}
\propto
\int d\bfa d\bfr_5 d\bfr_6e^{i\left(\bfa-\frac{\rho\bfk}{1+\rho}\right)
\cdot\bfr_5}e^{-i\left(\bfa+\frac{\bfk'}{1+\rho}-\bfk\right)\cdot\bfr_6}
e^{-\h\beta_1^2r_5^2}e^{-\h\beta_2^2r_6^2}
\qe
\eq
\propto
e^{i\frac{\bfk-\bfk'}{1+\rho}\cdot\bfr_5}e^{-\tbe^2r_5^2}
%=\beta_1^3\beta_2^3
%e^{-\frac{(\bfk-\bfk')^2}{(1+\rho)^24\tbe^2}
%\frac{\pi^{\frac{3}{2}}}{\tbe^3}(2\pi)^3
\propto
e^{-\frac{(\bfk-\bfk')^2}{4\tbe^2(1+\rho)^2}}.
\qe

For nucleon B the three momenta
$\bfb_1,\bfb_2,\ {\rm and}\ -\bfk-\bfb_1-\bfb_2$ 
give for the Jacoby coordinates
\eq
\bfp =\bfb_1-\bfb_2,\ \bfQ=\frac{\bfb_1+\bfb_2}{2},\ \ 
\bfq=-\bfk-\frac{3}{2}(\bfb_1+\bfb_2)
\qe
and hence the B nucleon wave function factor becomes
\eq
e^{-\frac{(\bfb_1-\bfb_2)^2}{4\alpha_1^2}} 
e^{-\frac{\left[\bfk+\frac{3}{2} 
(\bfb_1+\bfb_2)\right]^2}{3\alpha_1^2}}.
\qe

For the D nucleon the coordinates 
$ \bfb_1+\bfk-\bfk', \bfb_2, \ \ {\rm and}\ \ -\bfk-\bfb_1-\bfb_2
$
give for Jacoby coordinates
\eq
\bfp=\bfk-\bfk'+\bfb_1-\bfb_2, \bfQ=\frac{\bfk-\bfk'+\bfb_1+\bfb_2}{2},
\ \ {\rm and}\ \ \bfq=-\frac{3}{2}\bfk+\h\bfk'-\frac{3}{2}(\bfb_1+\bfb_2)
\qe
and the wave function becomes
\eq
e^{-\frac{(\bfk-\bfk'+\bfb_1-\bfb_2)^2}{4\alpha_2^2}}
e^{-\frac{\left[\frac{3}{2}\bfk-\h \bfk'+\frac{3}{2}(\bfb_1+\bfb_2)
\right]^2}{3\alpha_2^2}}.
\qe

Writing these factors as r-space transforms we have
\eq
\int 
d\bfb_1d\bfb_2d\bfr_1d\bfr_2d\bfr_3 d\bfr_4
e^{i(\bfb_1-\bfb_2)\cdot\bfr_1} 
e^{-i(\bfk-\bfk'+\bfb_1-\bfb_2)\cdot\bfr_2}
e^{-\tal^2(r_1^2+r_2^2)}
\qe
$$
\times e^{i[\bfk+\frac{3}{2}(\bfb_1+\bfb_2)]\cdot\bfr_3}
e^{-i[\frac{3}{2}\bfk-\h \bfk'+\frac{3}{2}(\bfb_1+\bfb_2)]
\cdot\bfr_4}e^{-\frac{3}{4}\tal^2(r_3^2+r_4^2)}.
$$

The integrals on $\bfb_1$ and $\bfb_2$ give the delta functions
\eq
\delta (\bfr_1-\bfr_2+\frac{3}{2}\bfr_3-\frac{3}{2}\bfr_4),
\delta (-\bfr_1+\bfr_2+\frac{3}{2}\bfr_3-\frac{3}{2}\bfr_4).
\qe
We use them to eliminate $\bfr_2$ and $\bfr_4$ to find
$\bfr_2=\bfr_1$ and $\bfr_4=\bfr_3$. 

The integral becomes
\eq
\int d\bfr_1 e^{-(\bfk-\bfk')\cdot\bfr_1}e^{-2\tal^2r_1^2}
\int d\bfr_3 e^{-i\h(\bfk-\bfk')\cdot \bfr_3} e^{-\frac{3}{2}
\tilde{\alpha}^2r_3^2}
\propto e^{-\frac{(\bfk-\bfk')^2}{8\tal^2}}
e^{-\frac{(\bfk-\bfk')^2}{24\tilde{\alpha}^2}}
\propto e^{-\frac{(\bfk-\bfk')^2}{6\tilde{\alpha}^2}}.
\qe
Combining with the result for the kaon we
have an exponent of
\eq
(\bfk-\bfk')^2\left[\frac{1}{4\tbe^2(1+\rho)^2}+\frac{1}{6\tal^2}
\right]
= k^2(1-\mu)\left[\frac{1}{2\tbe^2(1+\rho)^2}
+\frac{1}{3\tal^2}\right]
\qe
which agrees (in the limit of equal $\alpha$'s and $\beta$'s) with 
Eq. 41 of Ref. \cite{bs}. A numerical check was performed by doing
the integral on the momentum $\bfa$ and then doing the remaining
six-dimensional integral by quadrature.

\subsection{Diagram 2}

The integral corresponding to diagram 2 is
$$
\int d\bfb_1 d\bfc d\bfd_1 \phi_A\left(2\bfc+\frac{2\bfk}{1+\rho}
-2\bfk'
\right)\phi^*_C\left(2\bfc+\frac{2\rho\bfk'}{1+\rho}\right)
\phi_B\left(\bfb_1,\bfc,-\bfk-\bfb_1-\bfc\right)
$$
\eq \times
\phi^*_D(\bfd_1,\bfk-\bfk'+\bfb_1+\bfc-\bfd_1,-\bfk-\bfb_1-\bfc).
\label{dia2}\qe

Note that there is a typographical error in equation 35 of Ref. \cite{bs} 
where there is a minus sign in front of the term $2\bfk/(1+\rho)$ in the 
kaon A wave function \cite{swanson}. This is only a error of 
transcription; we agree with all of their final results.

For nucleon B the Jacoby variables are (we leave out $\bfQ$ which is not
needed)
$\bfp=\bfb_1-\bfc,\ \ \bfq=-\bfk-\frac{3}{2}(\bfb_1+\bfc)$
and for nucleon D they are
$
\bfp=2\bfd_1-\bfk+\bfk'-\bfb_1-\bfc,\ \ 
\bfq=-\frac{3}{2}\bfk+\h \bfk'-\frac{3}{2}(\bfb_1+\bfc).
$
With the vectors $\bfr_5$ and $\bfr_6$ being associated with the
kaon wave functions the imaginary part of the exponent arising 
from the Fourier transform is
\eq
(\bfb_1-\bfc)\cdot\bfr_1-[\bfk+\frac{3}{2}(\bfb_1+\bfc)]\cdot\bfr_2
-[2\bfd_1-\bfk+\bfk'-\bfb_1-\bfc]\cdot\bfr_3
+[\frac{3}{2}\bfk-\h \bfk'+\frac{3}{2}(\bfb_1+\bfc)]\cdot\bfr_4.
\qe
$$
+2\left(\bfc+\frac{\bfk}{1+\rho}-\bfk'\right)\cdot\bfr_5-
2\left(\bfc-\frac{\rho\bfk'}{1+\rho}\right)\cdot\bfr_6.
$$

The integrals on $\bfd_1$, $\bfb_1$ and $\bfc$ give $\bfr_3=0$, \
$\bfr_2=\frac{2}{3}\bfr_1+\bfr_4$, and $\bfr_6=\bfr_5-\bfr_1$ 
so that the integral becomes
\eq
\int d\bfr_1d\bfr_4d\bfr_5
e^{i[(-\frac{2}{3}\bfk-\frac{2\rho\bfk'}{1+\rho})\cdot\bfr_1+\h 
(\bfk-\bfk')\cdot\bfr_4+2\frac{(\bfk-\bfk')}{1+\rho}\cdot\bfr_5]} 
e^{-\alpha_1^2r_1^2}e^{-\frac{4}{3}\alpha_1^2(\frac{2}{3}\bfr_1+\bfr_4)^2}
e^{-\frac{3}{4}\alpha_2^2r_4^2}e^{-2\beta_1^2r_5^2}e^{-2\beta_2^2
(\bfr_5-\bfr_1)^2}.
\qe

Defining $\gamma^{\beta}_i=\beta_i^2/(\beta_1^2+\beta_2^2)$, we can
rewrite the $\bfr_5$ integral using Eq. \ref{b1} as
\eq
e^{2i\gamma^{\beta}_2(\frac{\bfk-\bfk'}{1+\rho})\cdot\bfr_1}
e^{-2\beta_{12}^2r_1^2}\int d\bfs 
e^{2i(\frac{\bfk-\bfk'}{1+\rho})\cdot\bfs}
e^{-4\tbe^2s^2}.
\qe
Now we do the integral on $\bfr_4$ using Eq. \ref{b1},
\eq
\int d\bfr_4 e^{-\h i(\bfk'-\bfk)
\cdot\bfr_4}e^{-\frac{3}{4}\alpha_2^2r_4^2}
e^{-\frac{3}{4}\alpha_1^2(\frac{2}{3}\bfr_1+\bfr_4)^2}
=e^{i\h(1-\gamma^{\alpha}_2)
(\bfk'-\bfk)\cdot\frac{2}{3}\bfr_1}e^{-\frac{3}{4}\alpha_{12}^2
\frac{4}{9}r_1^2}\int d\bft e^{-\h 
i(\bfk'-\bfk)\cdot\bft}e^{-\frac{3}{2}\tal^2t^2}.
\qe

The contribution of the $\bfs$ and $\bft$ integrals to the exponent is
\eq
\frac{4(\bfk-\bfk')^2}{(1+\rho)^2}\frac{1}{16\tbe^2}
+\frac{(\bfk-\bfk')^2}{4}\frac{2}{12\tal^2}
%\longrightarrow\frac{k^2}{12\alpha^2(1+\rho)^2}\left\{6g+(1+\rho)^2
%-\mu\left[(1+\rho)^2+6g\right]\right\}
=\frac{k^2}{2(1+\rho)^2\tbe^2}-\frac{\mu k^2}{2(1+\rho)^2\tbe^2}
+\frac{k^2}{12\tal^2}-\frac{\mu k^2}{12\tal^2}
\qe
so the contributions of the $\bfs$ and $\bft$ integrals to $a_2$ and $b_2$ 
are
\eq
\frac{1}{2(1+\rho)^2\tbe^2}+\frac{1}{12\tal^2};\ \ 
\frac{1}{2(1+\rho)^2\tbe^2}+\frac{1}{12\tal^2}.
\qe

The coefficient of $\bfr_1$ is now
\eq
-\frac{2}{3}\bfk-\frac{2\rho\bfk'}{1+\rho}+2\gamma^{\beta}_2\frac{\bfk-\bfk'}
{1+\rho}+\frac{1}{3}\gamma_1^{\alpha}(\bfk'-\bfk)
\qe
$$
=-\frac{[(2+\gamma_1^{\alpha})(1+\rho)-6\gamma^{\beta}_2]\bfk
+[-\gamma_1^{\alpha}(1+\rho)+6(\gamma^{\beta}_2+\rho)]\bfk'}
{3(1+\rho)}=C_2\bfk+C_2'\bfk'
$$
where
\eq
\gamma^{\alpha}_i=\frac{\alpha_i^2}{\alpha_1^2+\alpha_2^2}.
\qe

The Gaussian in the $\bfr_1$ integral becomes
\eq
e^{-(\alpha_1^2+2\beta_{12}^2+\frac{1}{3}\alpha_{12}^2)r_1^2}
\qe
so that the contribution of the $\bfr_1$ integral to the exponential
is 
\eq
\frac{(C_2\bfk+C_2'\bfk')^2}
{4(\alpha_1^2+2\beta_{12}^2+\frac{1}{3}\alpha_{12}^2)}
=k^2\frac{C_2^2+C_2'^2+2C_2 
C'_2\mu}{4(\alpha_1^2+2\beta_{12}^2+\frac{1}{3}\alpha_{12}^2)}.
\qe
so that the contributions to $a_2$ and $b_2$ are
\eq
\frac{C_2^2+C_2'^2}{4(\alpha_1^2+2 
\beta_{12}^2+\frac{1}{3}\alpha_{12}^2)}
;\ \ \ {\rm and}\ \ -\frac{C_2 
C_2'}{2(\alpha_1^2+2\beta_{12}^2+\frac{1}{3}\alpha_{12}^2)}.
\qe

A numerical check was done by first calculating the integral on
${\bf d}$ in the ``p'' factor of nucleon D. It can be integrated
on freely and gives only a contribution to the normalization of
$\pi^{\frac{3}{2}}\alpha_2^3$. The rest of the integral was done
by quadrature.

\subsection{Diagram 3}

The integral for diagram 3 is
$$
\int d\bfa d\bfb_2 d\bfc \phi_A\left(2\bfa-\frac{2\rho\bfk}{1+\rho}
\right)\phi^*_C\left(2\bfc+\frac{2\rho\bfk'}{1+\rho}\right)
\phi_B\left(\bfa-\bfk+\bfk',\bfb_2,-\bfa-\bfb_2-\bfk'\right)
$$
\eq \times
\phi^*_D(\bfa,\bfb_2,-\bfa-\bfb_2-\bfk').
\label{dia3}\qe
The Jacoby variables for nucleon B are: $
\bfp=-\bfk+\bfk'+\bfa-\bfb_2;\ \ \ 
\bfq=\h\bfk-\frac{3}{2}\bfk'-\frac{3}{2}(\bfa+\bfb_2)$ 
and for nucleon D:
$\bfp=\bfa-\bfb_2;\ \  \bfq=-\bfk'-\frac{3}{2} (\bfa+\bfb_2)$,
giving the integrand
\eq
e^{i\left\{(\bfa-\bfb_2-\bfk+\bfk')\cdot\bfr_1
+\h[\bfk-3\bfk'-3(\bfa+\bfb_2)]\cdot
\bfr_2-(\bfa-\bfb_2)]\cdot\bfr_3+[\bfk'+\frac{3}{2}(\bfa+\bfb_2)]\cdot\bfr_4
+\left(2\bfa-\frac{2\rho\bfk}{1+\rho}\right)\cdot\bfr_5-\left(2\bfc+
\frac{2\rho\bfk'}{1+\rho}\right)\cdot\bfr_6\right\}}
\qe
$$
\times 
e^{-\alpha_1^2r_1^2}e^{-\frac{3}{4}\alpha_1^2r_2^2}e^{-\alpha_2^2r_3^2}
e^{-\frac{3}{4}\alpha_2^2r_4^2}e^{-2\beta_1^2r_5^2}e^{-2\beta_2^2r_6^2}.
$$
Integration on the momenta gives $\bfr_6=0$, 
$\bfr_3=\bfr_1+\bfr_5$  and $\bfr_2=\bfr_4+\frac{2}{3}\bfr_5$.

With $\bfz\equiv \bfk'-\bfk$ we have
\eq
\int d\bfr_1d\bfr_4d\bfr_5 e^{i[\bfz\cdot\bfr_1+\h(\bfk-3\bfk')\cdot
(\bfr_4+\frac{2}{3}\bfr_5)+\bfk'\cdot\bfr_4-\frac{2\rho\bfk}{1+\rho}
\cdot \bfr_5]}
e^{-\alpha_1^2r_1^2}e^{-\frac{3}{4}\alpha_1^2(\bfr_4+
\frac{2}{3}\bfr_5)^2}e^{-\alpha_2^2
(\bfr_1+\bfr_5)^2}e^{-\frac{3}{4}\alpha_2^2r_4^2}e^{-2\beta_1^2r_5^2}
\qe
\eq
=\int d\bfr_1d\bfr_4d\bfr_5 e^{i\{\bfz\cdot\bfr_1-\h\bfz\cdot\bfr_4+
[\frac{1}{3}(\bfk-3\bfk')-\frac{2\rho\bfk}{1+\rho}]\cdot\bfr_5\}}
e^{-\alpha_1^2r_1^2}e^{-\alpha_2^2(\bfr_1+\bfr_5)^2}
e^{-\frac{3}{4}\alpha_2^2r_4^2}
e^{-\frac{3}{4}\alpha_1^2(\bfr_4+\frac{2}{3}\bfr_5)^2}e^{-2\beta_1^2r_5^2}
\qe
\eq
=\int d\bfr_5\int d\bft e^{i\bfz\cdot\bft}e^{-2\tal^2t^2}
e^{-i\gamma^{\alpha}_2\bfz\cdot\bfr_5}e^{-\alpha_{12}^2r_5^2}
\int d\bfs e^{-i\h\bfz\cdot\bfs}e^{-\frac{3}{2}\tal^2s^2}e^{
\frac{1}{3}\gamma_1^{\alpha}
i\bfz\cdot\bfr_5}e^{-\frac{1}{3}\alpha_{12}^2r_5^2}
\qe
$$\times
e^{i\left[\frac{1}{3}(\bfk-3\bfk')-\frac{2\rho\bfk}{1+\rho}\right]
\cdot\bfr_5}e^{-2\beta_1^2r_5^2}.
$$
The $\bfs$ and $\bft$ integrals give a contribution to the exponent
of  
\eq
\frac{z^2}{8\tal^2}+\frac{z^2}{24\tal^2}=\frac{z^2}{6\tal^2}
=\frac{k^2}{3\tal^2}-\frac{k^2\mu}{3\tal^2}.
\qe
With 
\eq C_3\equiv 
\frac{2[2\gamma^{\alpha}_2+(2\gamma^{\alpha}_2-3)\rho]}{3(1+\rho)}
;\ \ C_3'\equiv -\frac{2}{3}(2\gamma^{\alpha}_2+1)
\qe
the coefficient of $\bfr_5$ is
\eq
-\frac{4\gamma^{\alpha}_2-1}{3}\bfz
+\frac{\bfk-3\bfk'}{3}-\frac{2\rho\bfk}{1+\rho}
%=\frac{[1+2\gamma^{\alpha}_2+(2\gamma^{\alpha}_2-5)\rho]\bfk
%-(2\gamma^{\alpha}_2+3)(1+\rho)\bfk'}{3(1+\rho)}=
=C_3\bfk+C_3'\bfk'.
\qe
The $\bfr_5$ integral is now
\eq
\int d\bfr_5 e^{i(C_3\bfk+C_3'\bfk')\cdot\bfr_5}
e^{-(\frac{4}{3}\alpha_{12}^2+2\beta_1^2)r_5^2}
\qe
so its contribution to the exponent is
\eq
\frac{(C_3\bfk+C_3'\bfk')^2}{4(\frac{4}{3}\alpha_{12}^2+2\beta_1^2)}=
\frac{3k^2(C_3^2+C_3'^2+2C_3C_3'\mu)}{8(2\alpha_{12}^2+3\beta_1^2)}
\qe
and the contributions to $a_3$ and $b_3$ are
\eq
\frac{3(C_3^2+C_3'^2)}{8(2\alpha_{12}^2+3\beta_1^2)}\ \ {\rm 
and}\ \ 
-\frac{3C_3C_3'}{4(2\alpha_{12}^2+3\beta_1^2)}.
\qe

A numerical check was done by first doing the integral on
${\bf c}$ which only appears in kaon C. It can be integrated
on freely and gives a contribution to the normalization of
$(2\pi)^{\frac{3}{2}}\beta_2^3$. The remainder of the integral 
was done by quadrature.

\subsection{Diagram 4}

The integral for diagram 4 is

$$
\int d\bfa d\bfb_1 d\bfc \phi_A\left(2\bfa-\frac{2\rho\bfk}{1+\rho}
\right)\phi^*_C\left(2\bfc+\frac{2\rho\bfk'}{1+\rho}\right)
\phi_B\left(\bfb_1,\bfc,-\bfk-\bfb_1-\bfc\right)
$$
\eq \times
\phi^*_D(\bfk-\bfk'-\bfa+\bfb_1+\bfc,\bfa,\bfk-\bfb_1-\bfc).
\label{dia4}\qe

For the Jacoby variables for nucleon B we have: $
\bfp=\bfb_1-\bfc;\ \ \bfq=-\bfk-\frac{3}{2} (\bfb_1+\bfc)$
and for nucleon D: $\bfp=\bfk-\bfk'-2\bfa+\bfb_1+\bfc;\ \ 
\bfq=-\frac{3}{2}\bfk+\h\bfk'-\frac{3}{2}(\bfb_1+\bfc)$.

The imaginary part of the exponent is
\eq
i\left\{(\bfb_1-\bfc)\cdot\bfr_1-[\bfk+\frac{3}{2}(\bfb_1+\bfc)]\cdot\bfr_2
-(\bfk-\bfk'-2\bfa+\bfb_1+\bfc)\cdot\bfr_3
\right.
\qe
$$ \left.
-[-\frac{3}{2}\bfk+\h\bfk'-\frac{3}{2}(\bfb_1+\bfc)]\cdot\bfr_4
+\left(2\bfa-\frac{2\rho\bfk}{1+\rho}\right)\cdot\bfr_5
-\left(2\bfc-\frac{2\rho\bfk'}{1+\rho}\right)\cdot\bfr_6
\right\}.
$$

The $\bfa$, $\bfb_1$ and $\bfc$ integrals give $\bfr_3=-\bfr_5$, 
$\bfr_6=-\bfr_1$ and  $\bfr_2=\frac{2}{3}\bfr_1+\frac{2}{3}\bfr_5+\bfr_4$.

The Gaussian factors are
\eq
e^{-(\alpha_1^2+2\beta_2^2)r_1^2}e^{-(\alpha_2^2+2\beta_1^2)r_5^2}
e^{-\frac{3}{4}\alpha_2^2r_4^2}e^{-\frac{3}{4}\alpha_1^2
(\frac{2}{3}\bfr_1+\frac{2}{3}\bfr_5
+\bfr_4)^2}
\qe
and the imaginary part of the exponent
\eq
-2\left(\frac{1}{3}\bfk+\frac{\rho\bfk'}{1+\rho}\right)\cdot\bfr_1
+\h (\bfk-\bfk')\cdot\bfr_4+\left(\frac{1}{3}\bfk-\bfk'-
\frac{2\rho\bfk}{1+\rho}\right)\cdot\bfr_5\equiv\bfu_1\cdot\bfr_1+
\bfu_4\cdot\bfr_4+\bfu_5\cdot\bfr_5.
\qe

We make the transformation
\eq
\bfs=\bfr_1+\bfr_5;\ \ \bft=\bfr_1-\frac{w^2}{v^2}\bfr_5;\ \ 
\bfr_1=(v^2\bft+w^2\bfs)/(v^2+w^2);\ \ \bfr_5=v^2(\bfs-\bft)/(v^2+w^2)
\qe
where
\eq
v^2=\alpha_1^2+2\beta_2^2;\ \ w^2=\alpha_2^2+2\beta_1^2.
\qe
Now
\eq
v^2r_1^2+w^2r_5^2=\frac{v^2}{v^2+w^2}(v^2t^2+w^2s^2)
\qe
and the integral becomes
\eq
\int d\bfs d\bft d\bfr_4 e^{i\left[
\frac{v^2(\bfu_1-\bfu_5)\cdot\bft}{v^2+w^2}
+\frac{(w^2\bfu_1+v^2\bfu_5)\cdot\bfs}{v^2+w^2}
+\bfu_4\cdot\bfr_4\right]}
e^{-(v^2+w^2)t^2}e^{-\frac{v^2w^2s^2}{v^2+w^2}}
e^{-\frac{3}{4}\alpha_2^2r_4^2}e^{-\frac{3}{4}\alpha_1^2
(\frac{2}{3}\bfs+\bfr_4)^2}.
\qe
The integral on $\bft$ can now be done to give a contribution
to the exponent of
\eq
\frac{v^4(\bfu_1-\bfu_5)^2}{(v^2+w^2)^2}\frac{(v^2+w^2)}{4v^4}
=\frac{(\bfk'-\bfk)^2(1-\rho)^2}{8(\tal+2\tbe)(1+\rho)^2}.
\qe
The $\bfr_4$ integral becomes
\eq
\int d\bfr_4 
e^{i\bfu_4\cdot\bfr_4}e^{-\frac{3}{4}\alpha_2^2r_4^2}e^{-\frac{3}{4}
\alpha_1^2(\frac{2}{3}\bfs+\bfr_4)^2}
=e^{-i\frac{2}{3}\gamma_1^{\alpha}\bfu_4\cdot\bfs}
e^{-\frac{3}{4}\alpha_{12}^2(\frac{2}{3})^2s^2}
\int d\bfv e^{i\bfu_4\cdot\bfv} e^{-\frac{3}{2}\tal^2v^2}
\qe
and the $\bfv$ integral gives
\eq
\frac{(\bfk'-\bfk)^2}{4}\frac{2}{12\tal^2}=
\frac{(\bfk'-\bfk)^2}{24\tal^2}.
\qe
The $\bfs$ integral becomes
\eq
\int d\bfs e^{i\left[\frac{w^2\bfu_1+v^2\bfu_5}{v^2+w^2}-\frac{2}{3}
\gamma^{\alpha}_1\bfu_4\right]
\cdot\bfs}e^{-\left(\frac{v^2w^2}{v^2+w^2}+\frac{1}{3}
\alpha_{12}^2\right)s^2}
\qe
\eq
\frac{w^2\bfu_1+v^2\bfu_5}{v^2+w^2}-\frac{2}{3}\gamma_1^{\alpha}\bfu_4=
-\frac{\left[2w^2-v^2+\frac{6v^2\rho}{1+\rho}
+(v^2+w^2)\gamma_1^{\alpha}\right]\bfk
+\left[\frac{6b^2\rho}{1+\rho}+3v^2-(v^2+w^2)\gamma_1^{\alpha}\right]\bfk'}
{3(v^2+w^2)}
\qe
$$
=C_4\bfk+C_4'\bfk'.
$$
so the contribution to the exponent from the $\bfs$ integral is
\eq
\frac{(C_4\bfk+C_4'\bfk')^2}{4\left(\frac{v^2w^2}{v^2+w^2}+\frac{1}{3}
\alpha_{12}^2\right)}=\frac{k^2(C_4^2+C_4'^2+
2C_4C_4'\mu)}{4\left(\frac{v^2w^2}{v^2+w^2}+\frac{1}{3}
\alpha_{12}^2\right)}
\qe
and the contribution to $a_4$ and $b_4$ are
\eq
\frac{C_4^2+C_4'^2}{4\left(\frac{v^2w^2}{v^2+w^2}+\frac{1}{3}
\alpha_{12}^2\right)}\ \ {\rm and}\ \ 
-\frac{C_4C_4'}{2\left(\frac{v^2w^2}{v^2+w^2}+\frac{1}{3}
\alpha_{12}^2\right)}.
\qe   

For a numerical check, the $\bfa$ integral which only
affects kaon A and the ``$\bfp$'' factor of nucleon D was done. With
$\bfx=\bfa-\frac{\rho\bfk}{1+\rho}$ we have
\eq
\int d\bfa e^{-\frac{(\bfa-\frac{\rho \bfk}{1+\rho})^2}{2\beta_1^2}}
e^{-\frac{(\bfk-\bfk'-2\bfa+\bfb+\bfc)^2}{4\alpha_2^2}}
=\int d\bfx e^{-\frac{x^2}{2\beta_1^2}}
e^{-\frac{[\bfx+\frac{\rho\bfk}{1+\rho}-\h (\bfk-\bfk+\bfb
+\bfc)]^2}{\alpha_2^2}}
\qe
$$ 
=\frac{\sqrt{\pi}}{4}\left(\frac{2\alpha_2^2\beta_1^2}{
\alpha_2^2+2\beta_1^2}\right)^{\frac{3}{2}}
e^{\frac{\left[\frac{\rho\bfk}{1+\rho}-\h (\bfk-\bfk+\bfb
+\bfc)\right]^2}{\alpha_2^2+2\beta_1^2}}.
$$
With this first integral done, the remaining 6-dimensional integral 
was done numerically.

\end{document}